\documentclass[12pt]{article} 
\usepackage{graphicx}
\usepackage{amsmath,amssymb}

\begin{document}
\baselineskip 21pt
\setcounter{totalnumber}{8}

\bigskip

\centerline{\Large \bf Kinematics and Origin of Gas }
\centerline{\Large \bf in the Disk Galaxy NGC 2655}

\bigskip

\centerline{\large O.K. Silchenko$^1$, A.V. Moiseev$^{2,1}$, A.S. Gusev$^{1}$, and D.V. Kozlova$^3$}

\noindent
{\it Sternberg Astronomical Institute of the Lomonosov Moscow State University, Moscow, Russia}$^1$

\noindent
{\it Special Astrophysical Observatory of the Russian Academy of Sciences, Nizhnij Arkhyz, Russia}$^2$

\noindent
{\it Leibniz-Institut fur Astrophysik (AIP), Ander Sternwarte 16, 14482 Potsdam, Germany}$^3$

\vspace{2mm}
\sloppypar 
\vspace{2mm}

\bigskip

\noindent
The new observational data concerning distribution, excitation, and kinematics of the ionized gas
in the giant early-type disk galaxy NGC 2655 obtained at the 6m telescope of the Special Astrophysical Observatory
(SAO RAS) and at the 2.5m telescope of the Caucasian Mountain Observatory (CMO SAI MSU) are
presented in this work. The joint analysis of these and earlier spectral observations has allowed us to make a
conclusion about multiple nature of the gas in NGC 2655. Together with a proper large gaseous disk experiencing
regular circular rotation in the equatorial plane of the stellar potential of the galaxy for billions years,
we observe also remnants of a merged small satellite having striked the central part of NGC 2655 almost vertically
for some 10 million years ago.

\begin{center}

\end{center}

\clearpage

\section{INTRODUCTION}

The morphological type of lenticular galaxies was introduced by Hubble (1936) as a transitional type
between ellipticals and spirals. However, having a large-scale stellar disk in their structure, 
lenticulars did not show noticeable star formation in it, opposite to spirals. It is very early that
a hypothesis was proposed that star formation does not occur in the disks of lenticular galaxies,
because there is no gas there; and S0s have no gas, because it was somehow ''removed'', for example,
by interaction with hot intergalactic medium in a cluster (Gunn and Gott, 1972; Larson et al., 1980). 
However, since then the paradigm of the spiral (disk) galaxy evolution has changed since then, it became 
clear that the entire evolution of disk galaxies is governed by an inflow of cold gas from outside compensating
for any losses of it in the disk, in particular, the losses due to star formation (Tacconi et al., 2020).
Also, deep radio observations, both large-scale surveys like ALFALFA (Grossi et al., 2009) and targeted studies of
specific samples of early-type galaxies, such as the ATLAS-3D survey (Serra et al., 2012), showed that
almost half of the field lenticular galaxies has massive extended gaseous disks. Why does not the same star
formation take place in these disks as that in the disks of spiral galaxies?

Observations of gas kinematics in field lenticular galaxies have always shown impressive fraction of decoupled
rotation of gas and stars -- from 24\%\ (Kuijken et al., 1996) in earlier long-slit studies up to 36\%\ (Davis et al.,
2011) and even up to half of all lenticular galaxies in the extremely sparse environment (Katkov et al., 2015).
We have previously concluded (Silchenko et al. 2019) that the suppressed star formation in gas-rich lenticular
galaxies may be due to the off-plane inflow of the accretion flow: the gas falling into the potential well of the 
stellar disk suffers shocks, is heated, and becomes unable to form stars. We tested this hypothesis with a sample of
18 lenticular galaxies having extended gaseous disks, by observing them through panoramic spectroscopy, with the
Fabry-Perot scanning interferometer of the 6m SAO RAS telescope. We have constructed 2D line-of-sight velocity
distributions and have traced the orientation of the gas rotation plane in space along the galactic radius.
Indeed, star formation (particularly the star formation rings in lenticular galaxies) appeared to locate
only at the radii, where the gas lies onto the plane of the stellar disk. On the contrary, in inclined gaseous disks,star
formation does not proceed (Silchenko et al., 2019). One of the targets in our sample in this work was a nearby giant
lenticular galaxy NGC 2655. Figure~1 presents its images provided by the ground-based photometric
observations taken from the BASS survey and by high spatial resolution composite observations from the Hubble Space Telescope.

\begin{figure*}[!h]
\centerline{
 \includegraphics[width=16cm]{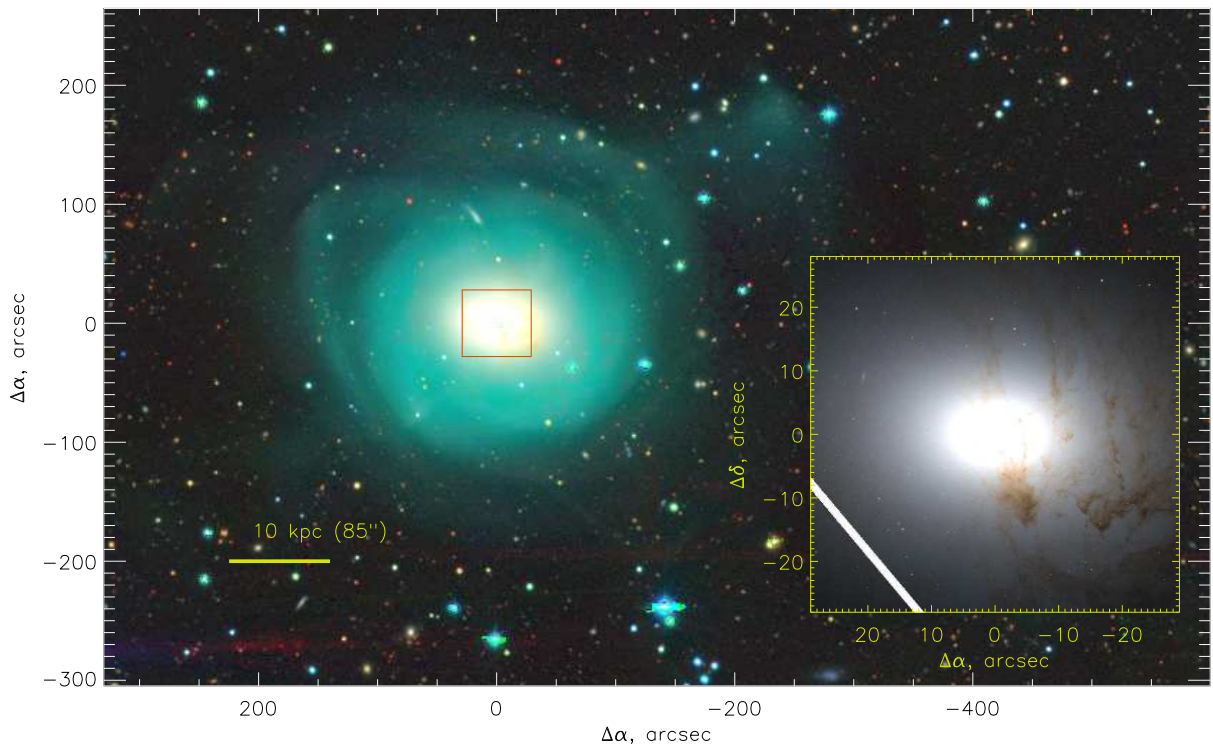}
 }
\caption{The images of NGC 2655 in composite colors: {\it the left plot} -- the deep broad-band image of the galaxy
taken from the DESI Legacy Imaging Surveys resource (Dey et al. 2019), 
{\it the right plot} --  the image of the central part of the galaxy obtained in broad-band filters by the Hubble Space
Telescope. At the {\it the right plot} one can see asymmetric dust rings produces by the projection of the circumnuclear polar
disk.}
\label{n2655im}
\end{figure*}

NGC 2655 is a giant disk galaxy in the center of a group: at a currently accepted distance to the galaxy of
24.4 Mpc (the scale is 118 pc/arcsec), its absolute magnitude is $M_K =-25$ (LEDA and NED), and the mass of the stellar 
population is $2 \cdot 10^{11}$ solar masses (Bouquin et al., 2018). The group includes seven galaxies brighter
than $M_B=-15$, all of them are of the late type (Garcia 1993). With this configuration, one would expect
that the whole gas content of NGC 2655 could result from accumulating the surrounding dwarfs by
the central galaxy. Indeed, NGC 2655 is abundant in neutral hydrogen: according to the earliest surveys, 
up to 3--6 billion solar masses of neutral hydrogen have been found in the galaxy (Lewis and Davies, 1973). 
It forms a giant disk with a diameter of five times the diameter of the stellar disk (Huchtmeier and Richter, 1982). 
The integrated star formation rate (SFR) estimated from the ultraviolet fluxes of the galaxy according to the data 
produced by the GALEX space telescope is 0.08 solar masses per year (Bouquin et al., 2018), which places the galaxy 
significantly below the Main Sequence classifying it as a ''galaxy with quenched star formation'' (Cortese et al. 2020).
At the same time, it should be noted that such SFR is anomalously low for the observed abundance of H I (Catinella et al. 2018). 
Detailed investigation of the spatial distribution of the neutral hydrogen density (Shane and Krumm 1983, Sparke et al. 2008) 
reveals the extension of the gaseous disk in a position angle of $\sim 110^{\circ}$; we are going to compare this orientation
with the parameters of the orientation of the stellar disk in Discussion in the present paper. As for the kinematics 
of the stars and gas, mapped for the central part of the galaxy through the panoramic spectroscopy, the gas demonstrates 
a polar rotation in the center with respect to the stars (Silchenko and Afanasiev, 2004; Dumas et al., 2007).

Another feature which is worth to be taken into account is the active nucleus of NGC 2655. Most researchers 
consider the nucleus of NGC 2655 as a Seyfert type II following our conclusion (Silchenko and Burenkov, 1990);
but, for example, Keel and Hummel (1988) noted a strong emission line [OI]$\lambda$6300 in the nucleus spectrum 
and classified it as a LINER. The NGC 2655 nucleus reveals a noticeable flux in X-ray including hard X-ray range 
(Terashima et al., 2002). High-resolution mapping of the NGC 2655 nucleus in the radio continuum detects a source 
with a steep spectrum which is compact both at wavelengths of 6 cm and 20 cm (Hummel et al., 1984); and from the
nucleus a jet comes out in the west-east direction, which curves farther from the nucleus to the north-south
direction (Ho and Ulvestad, 2001). Perhaps it is the jet that excites another compact radio source, at $15^{\prime \prime}$
(1.7 kpc) to the south-east from the nucleus, which demonstrates the same steep spectrum as the nucleus
(Keel and Hummel, 1988).

NGC 2655 is a testbed case of highly inclined rotation of gas in the absence of any star formation in a gas-rich S0,
which is of particular interest for us. However, the large-scale pattern of the velocity distribution in the extended gaseous
disk of NGC 2655 cannot be understood within a simple geometric model of a flat inclined rotation plane. Both the velocities 
and the brightness distribution of the emission lines in this galaxy reveal a very complex pattern. We have undertaken some
additional observations and are now ready to look into the details of how and when the gas has come to NGC 2655.

\section{NEW OBSERVATIONS}

We have already devoted several papers to the galaxy NGC 2655 (Silchenko and Burenkov, 1990;
Silchenko and Afanasiev, 2004; Silchenko et al., 2019), and we have since a tremendous collection of the spectroscopic
data obtained earlier with the 6-m BTA telescope. However, some incomprehensible moments remained in the interpretation 
of the ionized gas kinematics and, in order to clarify the whole picture, we decided to obtain additionally data.

\subsection{Mapping in Emission Lines}

We obtained an image of the galaxy with the NBI camera (Shatsky et al., 2020) in a narrow Halp filter
centered onto the complex of bright ionized-gas emission lines H$\alpha +$[NII]$\lambda \lambda$6548,6583,
having the transition peak at 656~nm, with the 2.5-m telescope of the Caucasus Mountain Observatory 
of SAI MSU (Shatsky et al., 2020) on January 10, 2018. The seeing during the observations was $2.5^{\prime \prime}$. 
The center wavelength of the filter used was 6560~\AA, the bandwidth was 77\AA, so both the [NII]$\lambda \lambda$6548,6583 
doublet lines and the hydrogen Balmer line H$\alpha$ fell there. At the same time, a feature of NGC~2655 is that the
[NII]$\lambda$6583 line is stronger than the H$\alpha$ line everywhere through the body of the galaxy (Silchenko and
Burenkov, 1990). The total exposure of the galaxy image obtained in the emission lines was 25 minutes. The image scale 
was 0.155$^{\prime \prime}$ per pixel. In addition to photometry in the narrow Halp filter, the galaxy was also
exposed in the neighboring continuum for 20 minutes (through the filter with a width of 100\AA\ centered on
6430\AA), so that after subtracting the image in the continuum from the image obtained in the Halp filter, it
would be possible to obtain a proper intensity distribution of the emission lines. Figure 2 shows the result of
this procedure together with the color map calculated from the broadband photometry in the BASS survey (taken from the
Legacy Survey resource, Dey et al. 2019). The morphology of the image in the emission lines represents a narrow loop, 
the center of which does not coincide with the center of the galaxy, plus a compact emission-line region to south-east from the
nucleus, which was previously detected in radiocontinuum emission (indicated in our picture as ESE). 
A red (dusty) loop outlines the inner edge of the gas emission loop and is especially bright to the south of the center of NGC 2655.
It is apparently associated with shock fronts generated by the collision of the polar nuclear gaseous disk with proper
gas of the galaxy, probably lying in the main plane of the galactic disk.

\begin{figure*}[!h]
\centerline{
 \includegraphics[width=6cm]{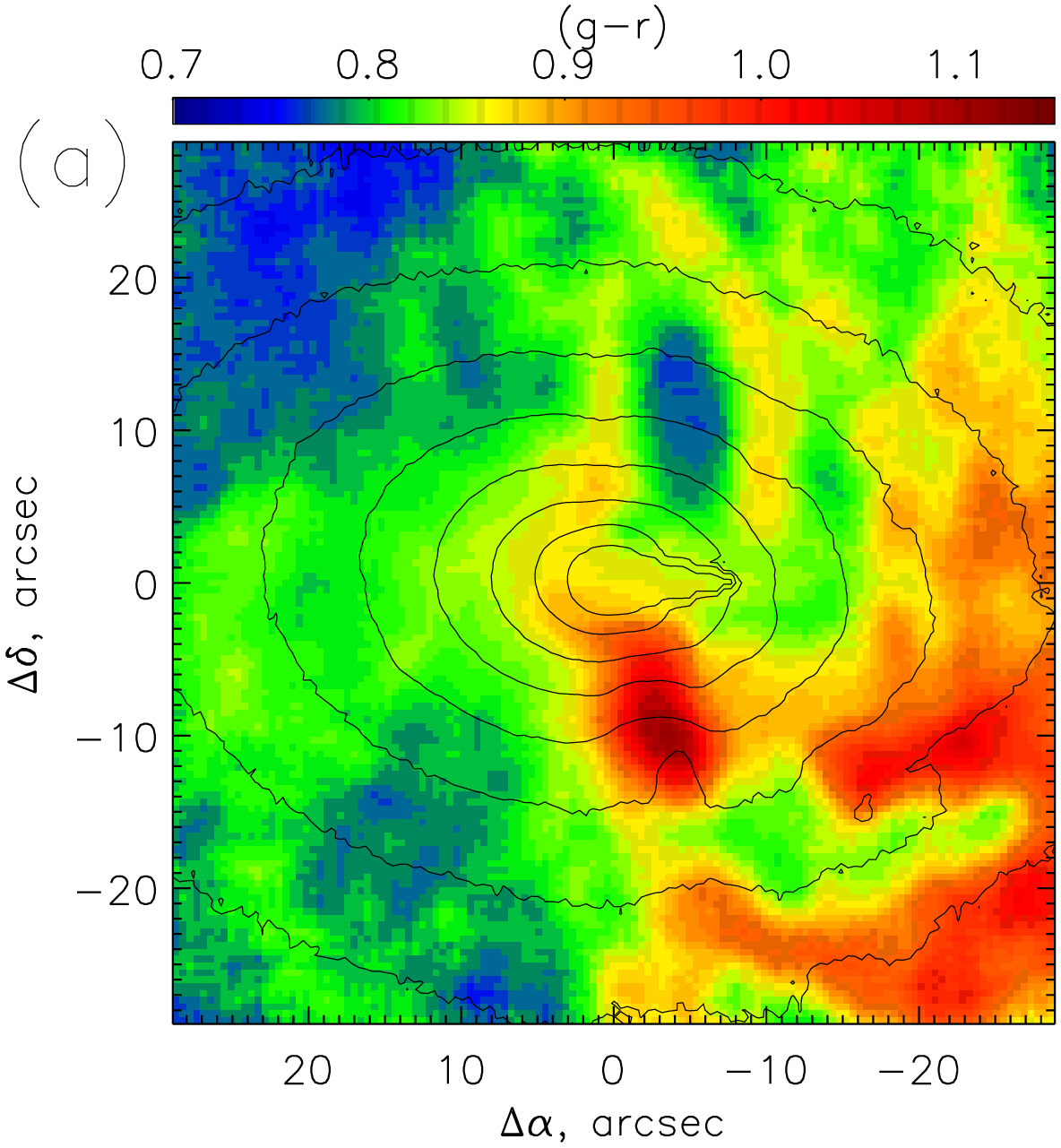}
 \includegraphics[width=6cm]{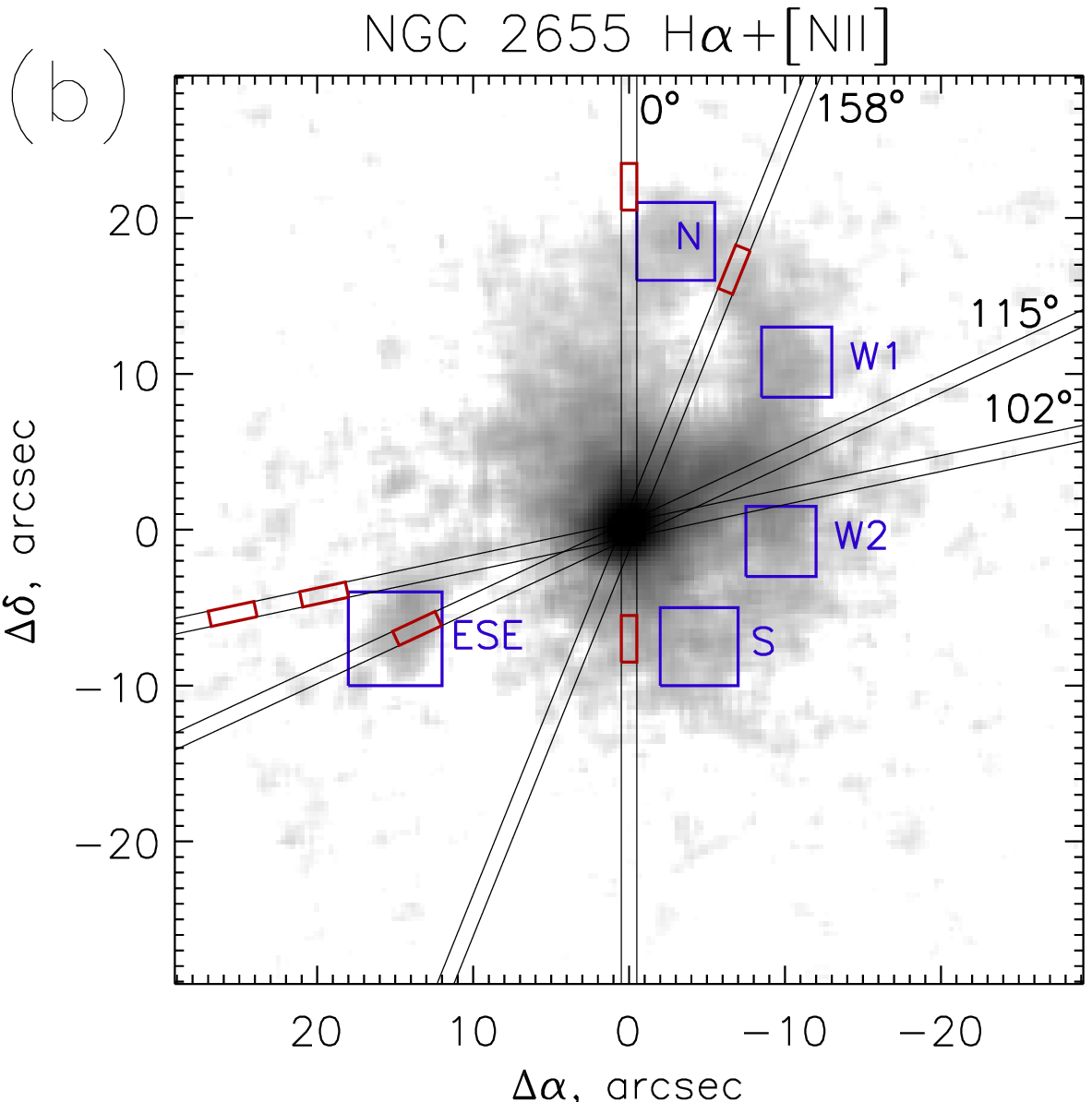}
 }
\caption{The central part of NGC 2655: {\it the left plot} -- the $g-r$ color image derived from the data of the BASS survey, 
{\it the right plot} --  the image through the narrow-band filter Halp, which includes 
the emission lines H$\alpha +$[NII]$\lambda \lambda$6548,6583, according to our data obtained at the 2.5m telescope of CMO SAI MSU,
after continuum subtracting. Some particular regions seen in the emissione lines are marked for further 
spectral analysis and discussion.}
\label{narrowim}
\end{figure*} 

\subsection{Long-slit Spectroscopy}

Additional long-slit spectroscopic data were obtained on May 26, 2022, at the BTA, the 6m SAO
RAS telescope, with the SCORPIO-2 multi-mode focal reducer (Afanasiev and Moiseev, 2011). The
VPHG1200@540 grism was used with a sensitivity maximum at 5400~\AA\ providing the full optical range
of spectroscopic observations in the wavelength range of 3650--7300~\AA\ with a resolution of about 5~\AA. The
slit was posed in two position angles: to include the ''radio loud'' ESE compact emission region (Fig. 2) and 
to catch the top of the northern part of the circumnuclear emission loop; the exposure times were 1600 sec and 800 sec 
respectively. The seeing quality during the spectroscopic observations in 2022 was $2.4^{\prime \prime}$. 
These long-slit cross-sections, together with the cross-sections at the position angles of $PA=102^{\circ}$ and $PA=0^{\circ}$,
previously obtained with the same instrument and the same grism, were used to measure the fluxes of various emission lines 
and their ratios for the selected regions at different distances from the center of the galaxy, and also to trace 
the line-of-sight velocities of the gas and the stars.

\section{EXCITATION OF THE IONIZED GAS}

Previously it was noted more than once (e.g., Sil'chenko and Burenkov 1990, Keel and Hummel 1988) that 
the strong emission lines in the spectrum of the NGC~2655 nucleus show flux ratios characteristic of 
a Seyfert type II active nucleus or a LINER one. Moreover, Keel and Hummel (1988), analyzing their spectrum 
of the ESE clump, modest as concerning the spectral range and the S/N ratio, have suspected that the spectrum of 
the ESE clump which is located in 1.8 kpc from the nucleus, is very similar to the nuclear spectrum 
in terms of the pattern of line flux ratios. Since the limitations on the energy of the nucleus did not allow 
explaining the ionized gas of the ESE clump as excited by the radiation of the central engine of the active nucleus, 
it was proposed that the gas excitation source here is a shock wave from the active nucleus jet which, according to 
radio interferometry, seems to be directed at the appropriate position angle. 

\begin{figure*}
\centerline{
\includegraphics[width=12cm]{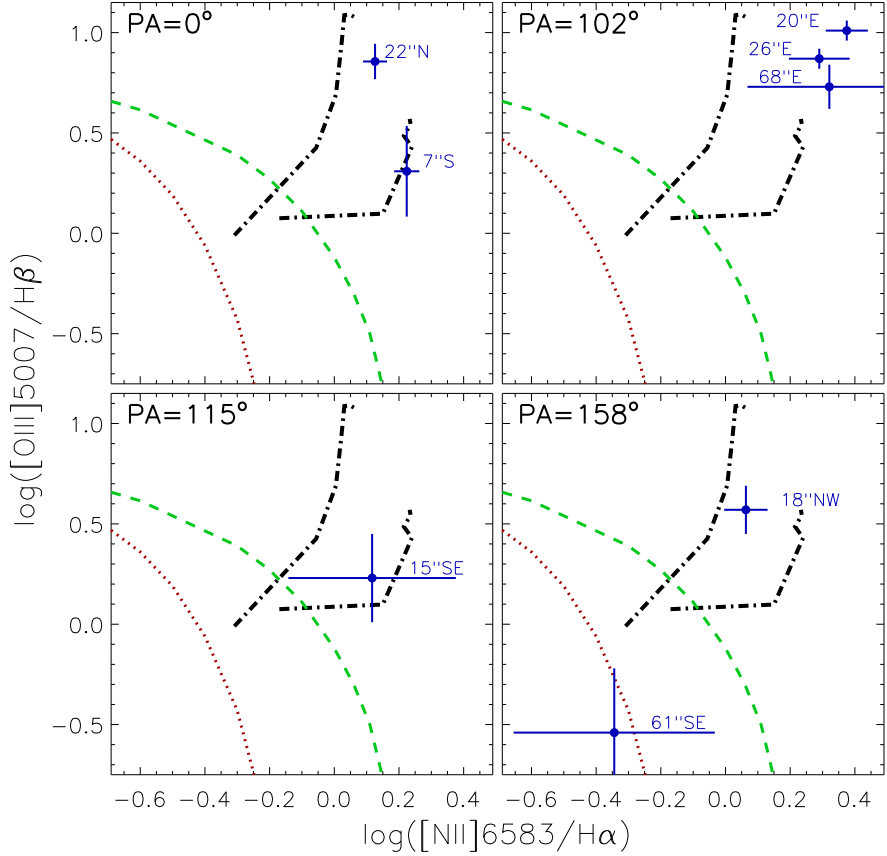}
 }
\caption{The diagnostic BPT-diagrams to determine a ionized-gas excitation source presented for four long-slit
cross-sections of NGC\,2655. The slit $PA$s are indicated in the upper left corner of each plot, the distance from the
center (and the direction along the slit, north or south, east or west) are given for every point (emission-line region).
The dotted red line and the green dashed line separate the areas for emission regions excited by young stars (to the left
and down) and other excitation mechanisms according to Kauffmann et al. (2003) and Kewley et al. (2001), respectively.
The dashed-dot fat lines show the models of shock excitation for the gas with solar metallicity and the typical electronic
density of $n=1$\,cm$^{-3}$ according to Allen et al. (2008). Along every model broken line the shock velocity rises
from bottom to top, from 200~km/s to 1000~km/s; the right broken line corresponds to the shock wave propagating in
low-density environment, and the left one -- to the shock model with precursor.}
\label{bptfull}
\end{figure*}

We obtained rather deep spectra with the 6-m BTA telescope at four different slit orientations. Measurements
of the flux ratios of the emission lines in these four spectra showed that the characteristic flux ratio dominated
by the highly-excited [OIII]$\lambda$5007 line is observed throughout the disk of NGC~2655, not only
in the ESE clump but also in the polar central loop (clumps N, W1, and S), and in the regular gaseous disk
of NGC~2655 up to the distance of 8~kpc from the center. Figure~3 shows the diagnostic diagrams -- the so called 
BPT-diagrams proposed by Baldwin et al. (1981) for diagnozing a gas ionization source --
to compare the ratio of the high-excitation [OIII]$\lambda$5007 line to the nearest Balmer hydrogen line H$\beta$, 
and the ratio of the low-excitation [NII]$\lambda$6583 line to the neighboring Balmer hydrogen line H$\alpha$,
for some selected areas of NGC~2655. The red dotted and green dashed lines separate the area of the emission regions 
excited by young stars (to the left and below the line) from other excitation mechanisms, according to the papers by
Kauffmann et al. (2003) and Kewley et al. (2001) respectively. Other excitation mechanisms are the ionization
either by the power-law spectrum of the active nucleus or by a shock wave: the BPT-diagram does not
makes it possible to distinguish between these two mechanisms. Since the regions under study are
located at different distances from the active nucleus, from 1 to 8 kpc, and the line ratios are similar for all
them, we think that we are dealing with gas excitation by a shock wave. Seven of the eight regions studied
contain the gas likely excited by shock wave. Although the areas of excitation by shock waves
and by a Seyfert nucleus overlap in the BPT-diagrams, in this case we are talking about the gas excitation at a
large distance from the center, and already Keel and Hummel (1988) have estimated that the radiation
from the active nucleus of NGC~2655 is not enough even to excite the ESE region at $15^{\prime \prime}$ from the
center, not to mention more distant regions. At the orientation $PA=102^{\circ}$ one can see how the shock wave 
slows down with distance from the center: if we compare the line ratios neasured by us with the Allen et al. (2008)
models, then from the point $r=20^{\prime \prime}$ to the point $r=60^{\prime \prime}$, the velocity of the shock 
wave falls down by some 150 km/s. Only a single region, at 7~kpc south of the
center in $PA=158^{\circ}$, is excited by young stars. This compact region is located at the periphery of the outer
disk and is also visible in the ultraviolet (Fig.~4). Since the gas in this region is ionized by young stars, we can
estimate its metallicity from the strong-line flux ratios calibrated using the HII region spectra modeled in detail.
We used two popular sources of such calibrations and obtained estimates for the oxygen abundance in the outer gas
for NGC~2655: $12+\log \mbox{(O/H)} =8.58\pm 0.18$~dex by the indicator N2 and $12+\log \mbox{(O/H)} =8.58\pm 0.16$~dex 
by the indicator O3N2 (Marino et al. 2013), or $12+\log \mbox{(O/H)} =8.71\pm 0.18$~dex by the indicator N2 and
$12+\log \mbox{(O/H)} =8.79\pm 0.21$~dex by the indicator O3N2 (Pettini and Pagel, 2004). Despite the low accuracy of
these estimates, we can still confirm that the metallicity of the gas is approximately solar, and this is at the
periphery of the galaxy disk!

\begin{figure*}
\centerline{
 \includegraphics[width=6cm]{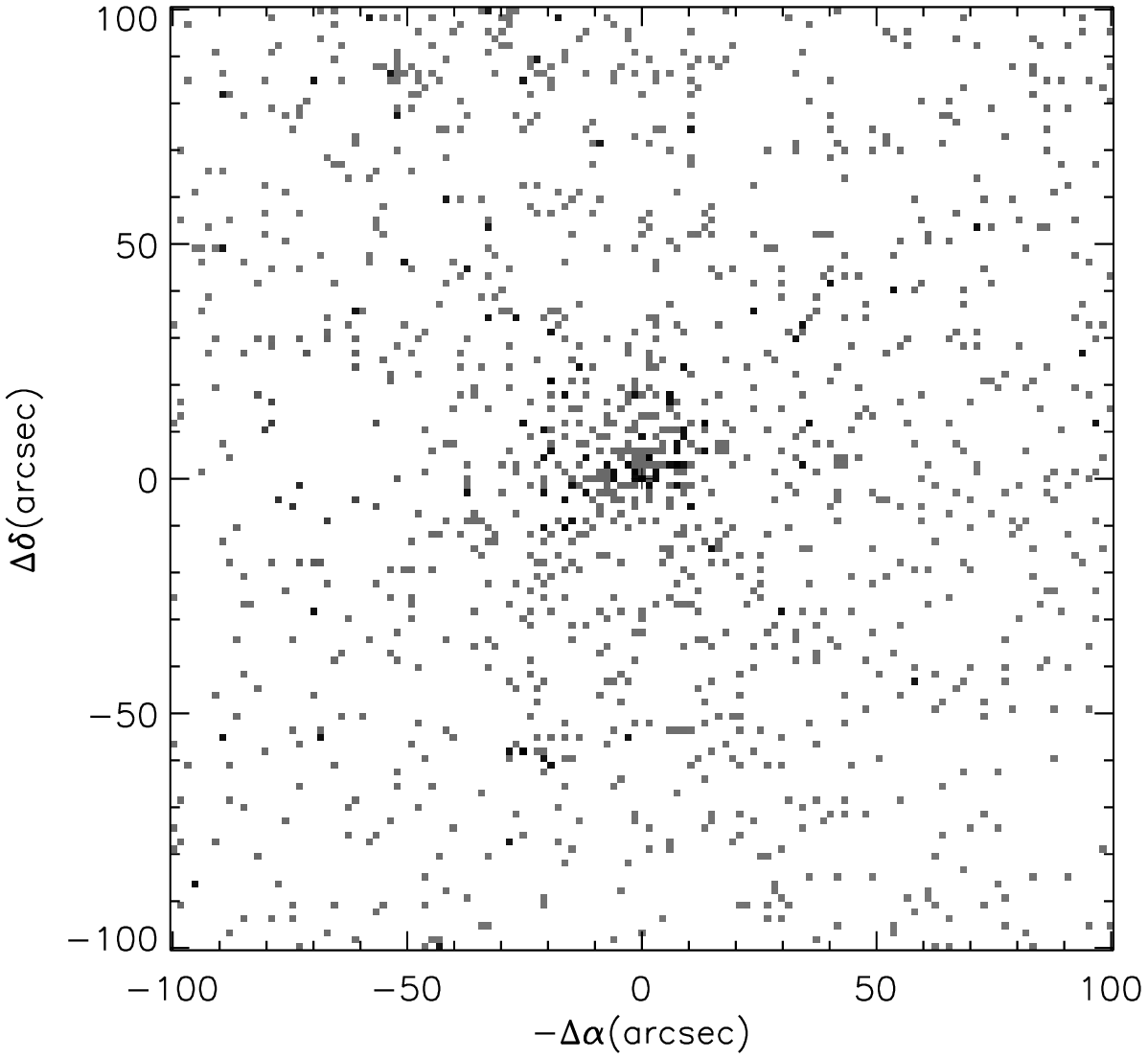}
 \includegraphics[width=6cm]{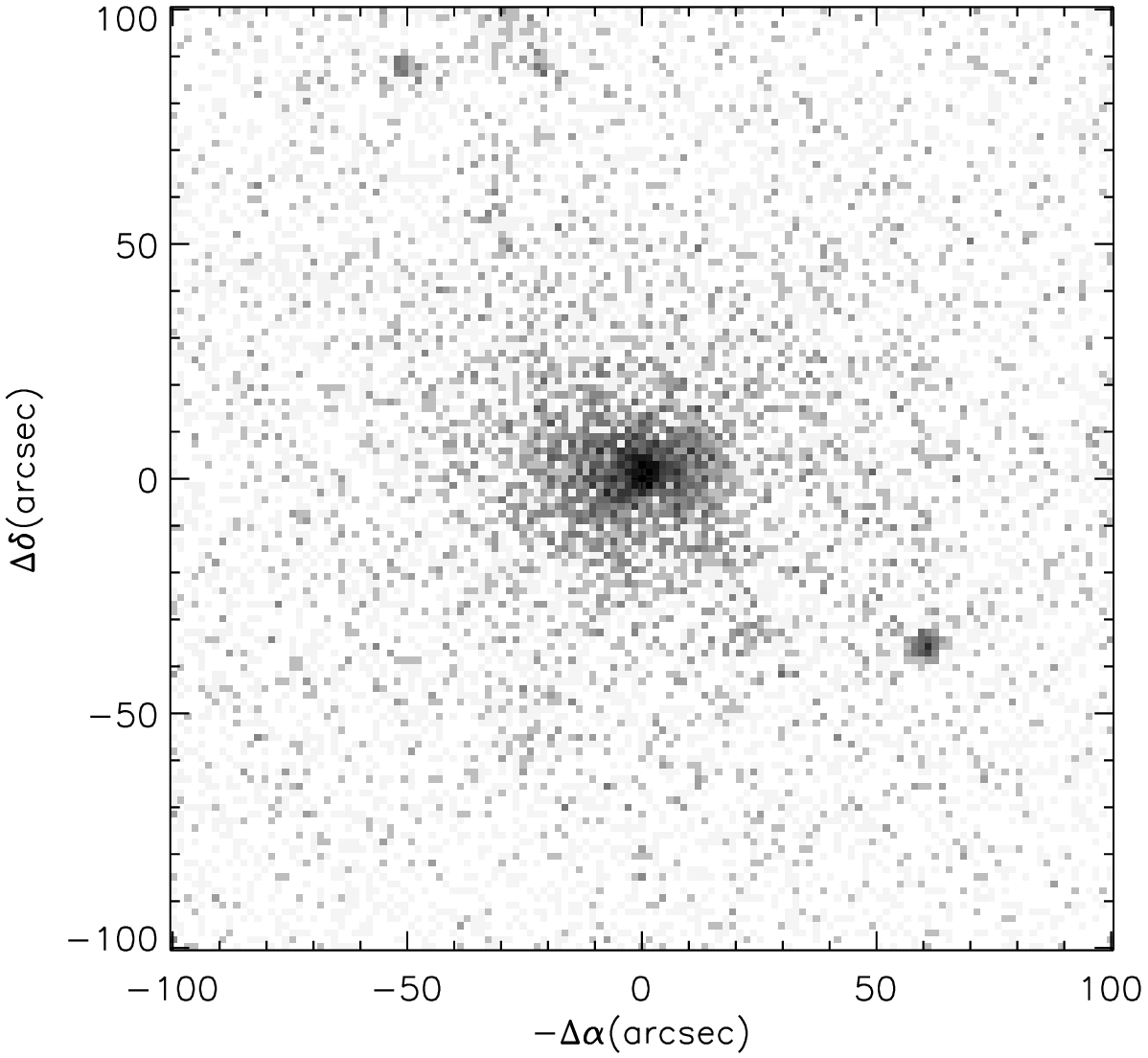}
 }
\caption{The ultraviolet maps of NGC\,2655 according to the GALEX data: {\it the left} -- the FUV map,
$\lambda \approx 1500$~\AA, {it the right} -- the NUV map, $\lambda \approx 2300$~\AA.}
\label{uv}
\end{figure*}

\section{THE DETAILED GAS KINEMATICS}

Earlier, we noted more than once the polar rotation of the ionized gas in the central region of NGC~2655
(Silchenko and Afanasiev, 2004; Silchenko et al., 2019). However the actual pattern of gas kinematics
throughout the entire galaxy can be much more complicated than simply warped rotation plane. The neutral 
hydrogen outside the stellar disk rotates regularly, in a circular manner according to the apparent orientation 
of the HI disk, with a kinematical major axis close to $PA=110^{\circ}$; Sparke et al. (2008) proposed a model 
with a smooth turn of the gaseous disk when going toward the center of the galaxy. Our data on the ionized gas 
in the outermost regions of the disk, at $R>40^{\prime \prime}$, also seem to agree with the stellar kinematics
(Silchenko et al., 2019). However, a lot of details in the distribution of the emission-line surface brightness
in Fig.~2 rather indicates not a smooth warp of the gaseous disk but the presence of several gas subsystems 
with different kinematics at the line of sight. This last hypothesis is also consistent with the shock excitation 
of the gas throughout the disk of NGC~2655.

\begin{figure*}
\centerline{
 \includegraphics[width=8cm]{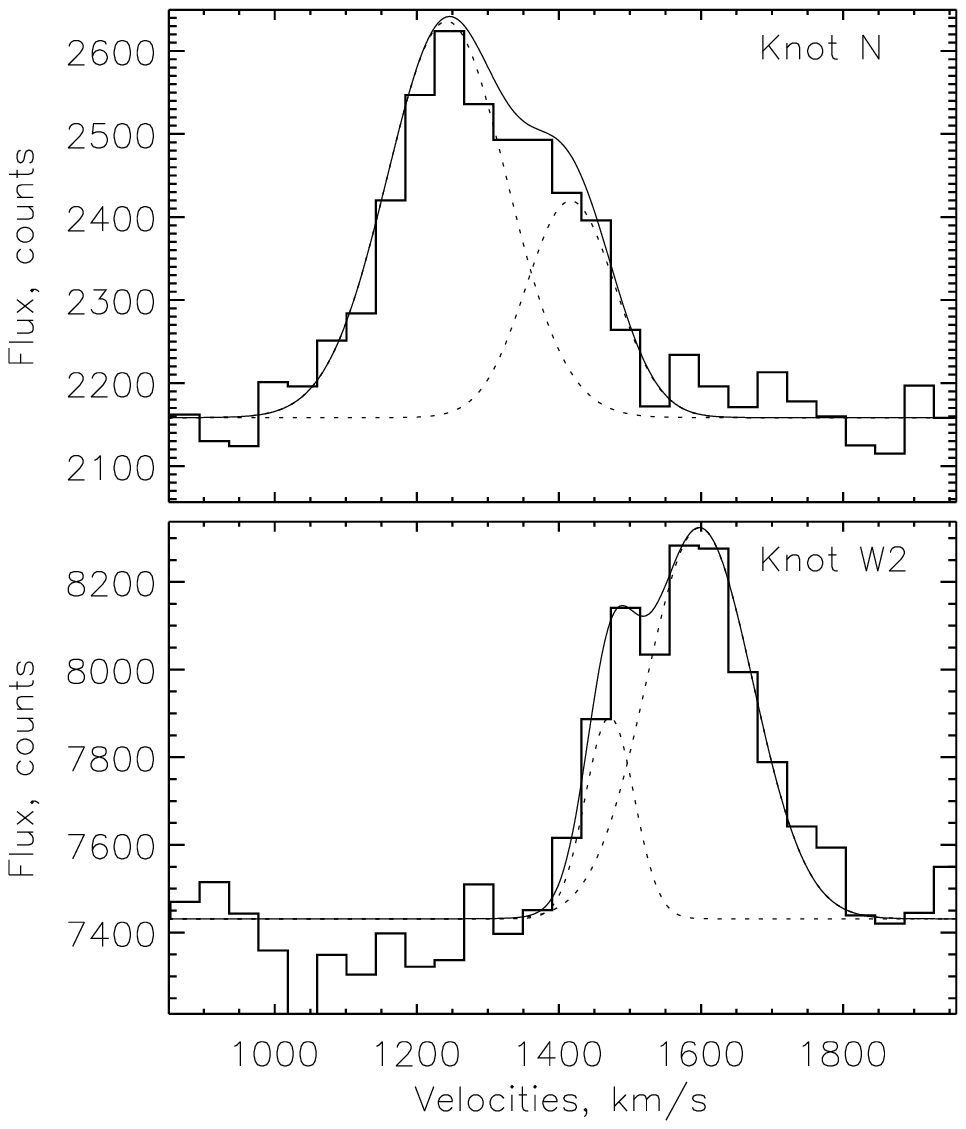}
 \includegraphics[width=8cm]{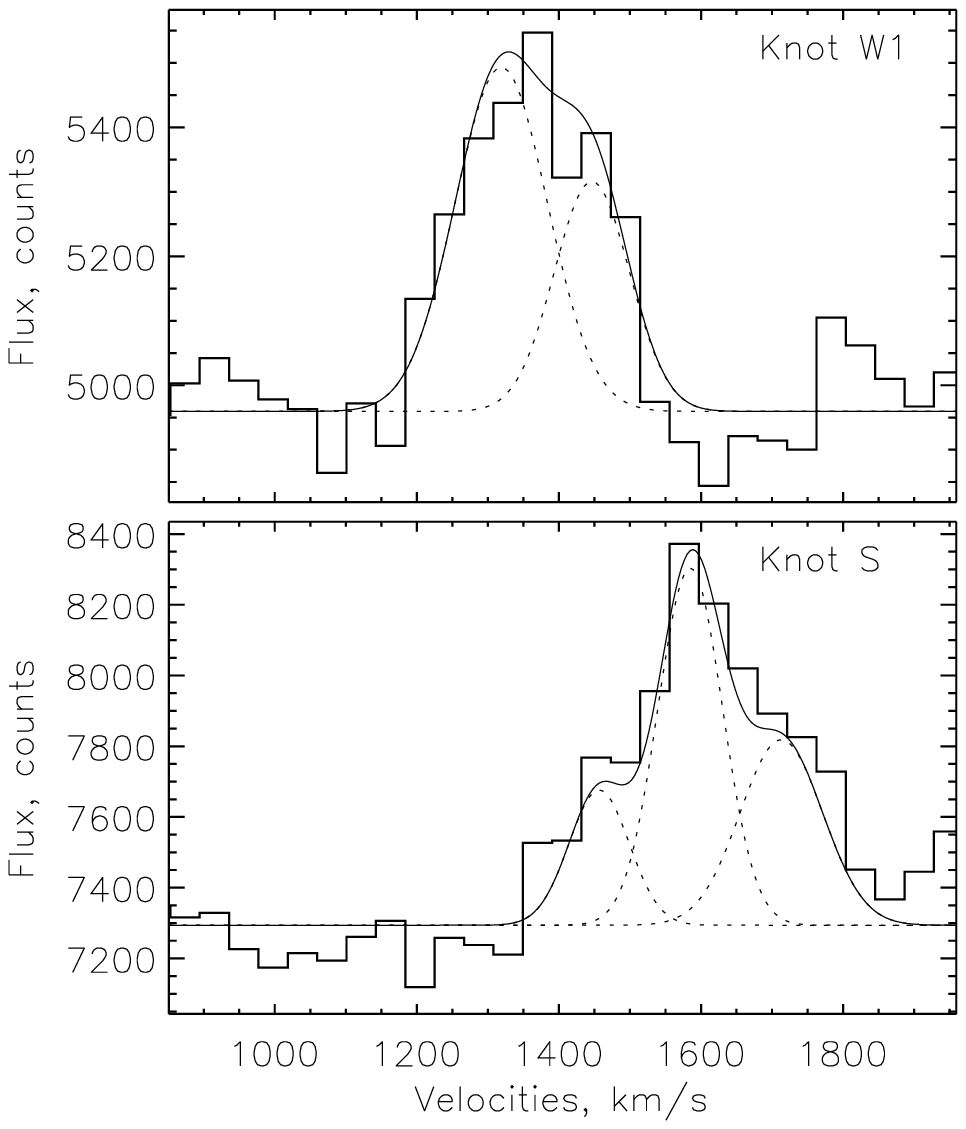}
 }
\caption{The Gauss-analysis of the [OIII]$\lambda$5007 emission-line profiles  for the compact emission-line regions
designated in Fig.~2 as  N (the upper left), W1 (the upper right), W2 (the bottom left), and S (the bottom right). 
The three former regions reveal the presence of two velocity components at the line of sight each:
$1196\pm 12$~km/s and $1371 \pm 16$~km/s (N), $1273\pm 33$~km/s and $1401 \pm 37$~km/s (W1), 
$1560\pm 6$~km/s and $1431 \pm 6$~km/s (W2), respectively. 
The southern loop clump reveals three velocity components, $1539\pm 11$~km/s, $1667\pm 32$~km/s, and $1371 \pm 16$~km/s.}
\label{gauss_fp}
\end{figure*} 

Using the benefit of rather high spectral resolution of our data obtained with the Fabry-Perot scanning
interferometer, as a second step of our analysis of these data, we decided to take a closer look at the line profiles;
the line analyzed is the [OIII]$\lambda$5007 emission line scanned in the narrow spectral range over the entire body 
of NGC~2655 with the Fabry-Perot interferometer (FPI). The line profiles appeared to be complex and multi-component.
Figure~5 presents the examples of the Gaussian line fitting for the loop areas marked as N, W1, W2, and S in Fig.~2. 
Let us note that although the FPI instrumental profile differs from the pure Gaussian one and can be rather described 
by a Voigt profile, but in the case of the given FPI, the observed line profiles differ little from the Gaussian one 
which can be clearly seen in Fig. 5.2. In every region N, W1, W2, and S, we can distinguish at least two components 
with different line-of-sight velocities. In the N and S regions, the stronger components imply the polar
rotation of the loop; but there are also weaker components demonstrating line-of-sight velocities
close to the systemic velocity of the galaxy, 1400 km/s, that is expected for the gas at the minor axis of the disk. 
Obviously, the weak components belong to the gaseous disk rotating in the plane of the stellar disk, whose isophote
major axis (and the line of nodes) is aligned close to the west-east direction.

\begin{figure*}
\centerline{
 \includegraphics[width=16cm]{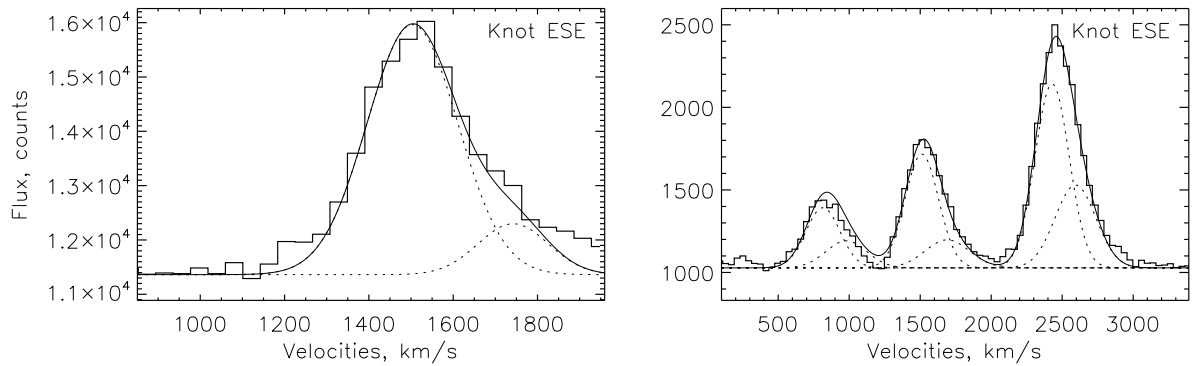}
 }
\caption{The Gauss-analysis of the emission-line profiles for the clump ESE: the [OIII]$\lambda$5007 line has
the velocity components $1499\pm 99$~km/s and $1737 \pm 377$~km/s, according to the Fabry-Perot data analysis;
and the long-slit data gives $1517\pm 35$~km/s and $1730 \pm 295$~km/s for the H$\alpha$ profile, and 
$1490\pm 36$~km/s and $1677 \pm 138$~km/s for the nitrogen doublet profile.}
\label{gauss_ese}
\end{figure*}

For the radio-loud ESE emission region located at 1.8 kpc southeast of the nucleus, Fig.~6 shows the results
of the Gaussian fitting for three lines: for the oxygen line derived from the Fabry-Perot data and for the hydrogen
H$\alpha$ and the nitrogen doublet according to the long-slit spectroscopy data. Although the weak component is measured
here with low accuracy, but for all three elements it occurs that in the ESE region there is the gas with a line-of-sight
velocity of about 1700~km/s; it is larger by 300~km/s than the systemic velocity of the galaxy. The gas with the similar
velocity is observed at the southwestern edge of the gaseous disk according to the Fabry-Perot velocity field
(Silchenko et al., 2019), and this velocity does not match any circular rotation models. Apparently,
as regarding the ESE clump, this may be a compact remnant of a satellite which has hit the NGC~2655 disk 
with a high impact velocity of 400--500~km/s almost at a right angle to the stellar disk. The whole configuration
with a destroyed companion and a polar circumnuclear loop looks like the destroyed Milky Way companion dSgr stretched 
into a polar stream in our Galaxy (Ibata et al., 2001; Laporte et al., 2018). And the NGC~2655 proper gas, which was hit 
in the center by the fallen companion, should have lost momentum in the shock wave and inflow into the nucleus; 
perhaps, this is what fuelled the current activity of the nucleus.

\section{DISCUSSION}

\subsection{Structure and Stellar Kinematics of NGC 2655}

NGC~2655 is a giant early-type disk galaxy. It is commonly accepted that such galaxies should have a very
large dominant bulge. Indeed, a detailed morphological analysis and decomposition of the galaxy image into components
undertaken as a part of the S4G survey of galaxies (Sheth et al., 2010) showed that the disk contributes no more 
than 42\%\ to the near-infrared luminosity -- and then to the stellar mass (Salo et al., 2015).
According to this decomposition, the exponential disk starts to dominate in the surface brightness at the
radii of $R>50^{\prime \prime}$, while closer to the center, the surface brightness profile represents a combined 
contribution of the bulge and bar. Why have the Salo et al. (2015) team decided that NGC~2655 has a bar, even
though the galaxy is not classified as SB in any catalog? This is due to the fact that the major-axis orientation
of the isophotes of the inner components -- the one with the $PA_1=82^{\circ}$, and the other with $PA_2=85.6^{\circ}$, --
differ from the orientation of the outermost disk isophotes, $PA_0=110^{\circ}$, which is commonly treated as the
orientation of the line of nodes (under the assumption of the round intrinsic shape of the disk). As a result, the 
NGC~2655 image deprojection undertaken in the S4G survey by the Salo et al. (2015) team exactly with this line-of-nodes
orientation, $PA=110^{\circ}$, has given the intrinsic galaxy shape with the oval inner components; the
Salo et al. (2015) team considered one of them as a triaxial bulge, and the other as a bar.

We do not agree with this interpretation of the structure of NGC~2655. The fact is that the stellar LOS velocity
field obtained for the central part of the galaxy with the SAURON IFU looks like a regular circular rotation
(Dumas et al., 2007). We analyzed this velocity field using the tilted ring method and found
the line-of-nods orientation for the stellar-component rotation plane $PA=263^{\circ} \pm 3^{\circ}$ up 
to the distance of $25^{\prime \prime}$ from the center. Dumas et al. (2007) obtained with the kinemetry method 
$PA=266^{\circ} \pm 1^{\circ}$ using the same data. The exact coincidence of the orientations
of the photometric and kinematical major axes proves that the stars in the center of NGC~2655 rotate 
in circular orbits within the axisymmetric potential: the galaxy has no bar.

\begin{figure*}[t]
\centerline{
 \includegraphics[width=5cm]{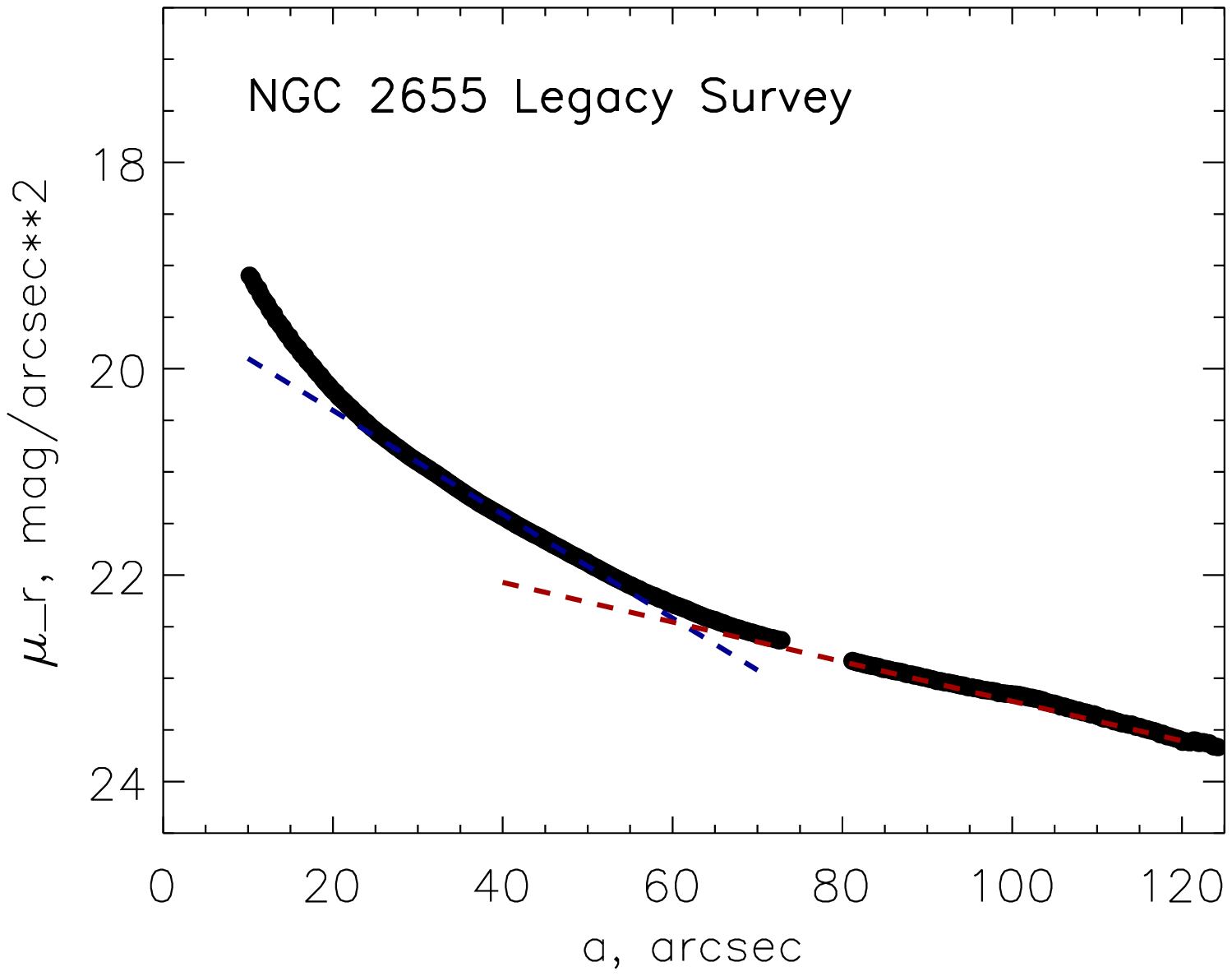}
 \includegraphics[width=7cm]{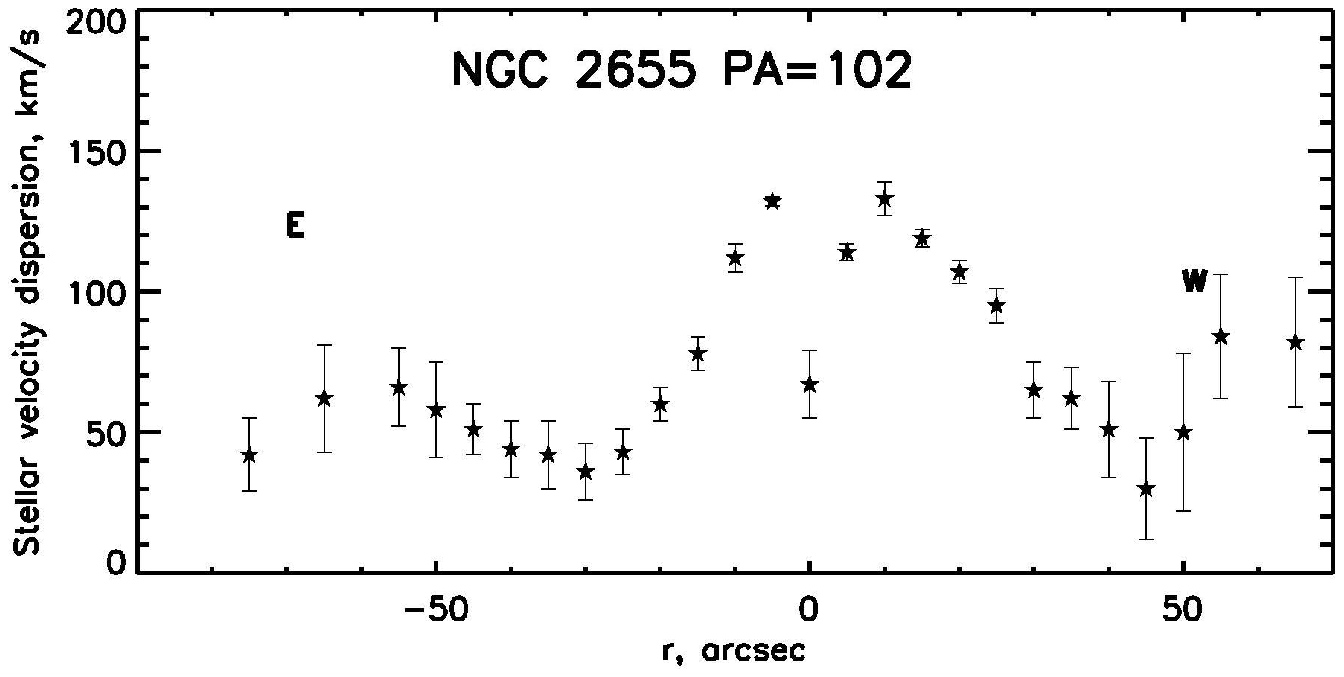}
 }
\caption{The azimuthally-averaged surface-brightness profile of NGC\,2655, according to the BASS survey data (taken from
Legacy Survey, Dey et al. 2019), 
fitted by two exponential segments, $\mu _r =19.4 + 1.086 R^{\prime \prime}/21.6^{\prime \prime}$, in the radius range 
of $R=26^{\prime \prime} - 50^{\prime \prime}$, and $\mu _r =21.3 + 1.086 R^{\prime \prime}/56.5^{\prime \prime}$, in the
radius range of $R=70^{\prime \prime} - 120^{\prime \prime}$ ({\it the left plot}), and the LOS stellar velocity dispersion 
profile in the $PA=102^{\circ}$ cross-section, according to the long-slit data ({\it right}).
}
\label{disk}                                                               
\end{figure*}

Another important diagnostic feature of a thin stellar disk is that it must be dynamically cold: 
its rotation velocity must be several times greater than the stellar velocity dispersion.
Figure~7, right, shows the profile of stellar velocity dispersion that we measured along the cross section
with a long slit at $PA=102^{\circ}$. The stellar line-of-sight velocities and velocity dispersions were
measured by the cross-correlation method similar to that we used in the paper by Silchenko et al. (2019).
Already at the radius of $R=30^{\prime \prime}$, the stellar velocity dispersion drops to 50~km/s: this 
is the radial boundary, where the thin stellar disk begins to dominate. We have also shown in Fig.~7, left,
the decomposition of the surface brightness profile consistent with the dominance of the disk so
close to the center: the photometric disk of NGC~2655 has the type III profile, that is, it consists 
of two exponential segments, the inner one with a smaller scalelength than the outer (which was also found in the S4G survey).

\begin{figure*}
\centerline{
\includegraphics[width=8cm]{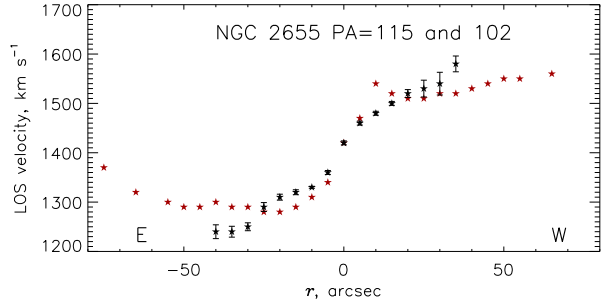}
}
\caption{The stellar LOS velocity profiles along two long-slit cross-sections close to the photometric major axis: 
the red stars are for $PA=102^{\circ}$, and the black stars -- for $PA=115^{\circ}$.}
\label{longslit_maj}
\end{figure*}

Thus, we can state that NGC~2655 has two exponential disks: they have different scalelengths, but they
also have different orientations of the isophote major axis. And along with this, the orientation of the major
axis of the inner isophote is supported by the stellar kinematics: as the analysis of the two-dimensional velocity
field shows, this is indeed the line of nodes of the circular rotation plane. As for the outer disk, for which
the orientation of the photometric major axis $PA=110^{\circ}$ was found in the S4G survey, here we cannot
properly compare it with the orientation of the kinematical major axis: there is no two-dimensional stellar
velocity field over a large extension, $R>20^{\prime \prime}$, of the galaxy disk. But we have long-slit 
cross-sections in different slit orientations. Figure~8 compares the line-of-sight velocity profiles for 
the stellar component in the slit orientations $PA=102^{\circ}$ and $PA=115^{\circ}$. We can note
that at the radius of $R=40^{\prime \prime}$ the rotation velocity projected onto the line of sight in $PA=115^{\circ}$ 
is larger than that in $PA=102^{\circ}$. This means that the kinematical major axis in the outer stellar disk 
is closer to $PA=115^{\circ}$ than to $PA=102^{\circ}$ which excludes the line-of-nodes orientation found for 
the inner stellar component. At the same time, the photometric major-axis orientation of the outer disk
may be the orientation of the line of nodes: our kinematical cross-sections with a long slit do not exclude
this. It appears that the internal and external rotation axes of the stellar disk of NGC~2655 are inclined to
each other, in other words, NGC~2655 is a multi-spin galaxy.

\subsection{The Origin of Gas in NGC 2655}

The orientations of the huge disk of neutral hydrogen and the outer stellar disk in NGC~2655 coincide
with each other both spatially and kinematically. Previously, Sparke et al. (2008) noted that two billion
solar masses of cold gas is too much for one minor merger, and several such events are needed (but with the same
orientation of the infall orbits, because all the gas rotates in the same plane). Now we understand that
these multiple minor mergers should have brought not only several billion solar masses of gas, but also several
billion solar masses of stars for the outer stellar disk of NGC~2655, which makes the supposed multiple
minor merging a quite incredible event. Opposite to Sparke et al. (2008), we conclude that the outer gaseous 
disk lies within the outer stellar disk, and even current star formation is taking place somewhere in it: 
it is at the southern edge of the disk that we have detected the gas emission excited by young stars, and
the northern arc shows also an excess of ultraviolet (Fig.~4). The metallicity of the gas in this outer disk is
solar, which is atypical for dwarf galaxies that Sparke et al. (2008) have suggested as the source of
NGC~2655 gaseous disk. The entire external configuration of the galaxy resembles a classical large disk of a
spiral galaxy, which, according to modern evolution concepts, is cumulated over billions years by smooth external
accretion of cold gas (Tacconi et al., 2020), albeit from a source undefined still at a global scale.

But minor merging certainly took place in NGC~2655. It also brought along a noticeable amount of the gas with the spin 
strongly decoupled from the regular rotation of the outer disk, both stellar and gaseous. Apparently, a
companion fell onto the galaxy almost vertically, and now, within two kiloparsecs from the center, we
observe the remnants of the destroyed companion as a circumpolar loop -- the picture is very similar 
to Sagittarius dwarf torn apart by the Milky Way. But in the case of NGC~2655, there was much more gas 
in the merged companion. The gas of the vertically infalling companion hit the gaseous disk
of NGC~2655 being in regular rotation, and this collision inevitably resulted in the development of shock
fronts. The shock wave has not only excited the gas in the polar loop, it ran outward across the large galactic
gaseous disk. At distances of up to 8 kpc from the center, we observe the gas of the large disk excited by this
shock wave, although, the kinematics of this gas to the east of the nucleus is little affected and exhibits rotation
consistent with that of the stellar disk. If the shock wave propagated at an average velocity of 1000 km/s ,
then the impact could have taken place approximately $10^7$ years ago.

\section{ACKNOWLEDGMENTS}

The paper is based on the observational data obtained with the 6-m telescope at the Special
Astrophysical Observatory of the Russian Academy of Sciences (BTA SAO RAS) and with the 2.5-m telescope
at the Caucasus Mountain Observatory of the Sternberg Astronomical Institute of the Moscow State University.
The spectroscopic analysis was supported by the grant of the Russian Science Foundation no.22-12-00080,
and the narrow-band photometry -- by the grant of the Russian Fund for Basic Research no. 20-02-00080. The
observations at the BTA SAO RAS telescope are supported by the Ministry of Science and Higher
Education of the Russian Federation; the observational technique is improved in the frame of the
National project "Science and universities". In our analysis we used data from publicly accessible archives and databases:
the Lyon-Meudon Extragalactic Database (LEDA) maintained by the LEDA team at the Lyon Observatory
CRAL (France) and the NASA/IPAC Extragalactic Database (NED) operated by the Jet Propulsion Laboratory
of the California Institute of Technology under contract with the National Aeronautics and Space Administration 
(USA). We also invoked the data from the GALEX space telescope for our analysis. The NASA GALEX data were
taken from the Mikulski Archive for Space Telescopes (MAST). In our figures we have also used the plots provided
by observations made with the NASA/ESA Hubble Space Telescope and obtained from the Hubble Legacy Archive,
which is a collaboration between the Space Telescope Science Institute (STScI/NASA),
the Space Telescope European Coordinating Facility (ST-ECF/ESA) and the Canadian Astronomy Data Centre (CADC/NRC/CSA).
The broad-band photometry is based on the data taken from the Legacy Survey resource (the BASS survey).
The Legacy Surveys consist of three individual and complementary projects: the Dark Energy Camera Legacy Survey
(DECaLS; Proposal ID no.2014B-0404; PIs: David Schlegel and Arjun Dey), the Beijing-Arizona Sky Survey
(BASS; NOAO Prop. ID no.2015A-0801; PIs: Zhou Xu and Xiaohui Fan), and the Mayall z-band Legacy Survey
(MzLS; Prop. ID no.2016A-0453; PI: Arjun Dey).
BASS is a key project of the Telescope Access Program (TAP), which has been funded by the National Astronomical
Observatories of China, the Chinese Academy of Sciences (the Strategic Priority Research Programs,
Grant no. XDB09000000), and the Special Fund for Astronomy from the Ministry of Finance.
The BASS is also supported by the External Cooperation Program of Chinese Academy of Sciences (Grant no. 114A11KYSB20160057),
and Chinese National Natural Science Foundation (Grant no. 12120101003, no. 11433005).

\end{document}